\documentclass{article}

\usepackage{microtype}
\usepackage{graphicx}
\usepackage{subfigure}
\usepackage{xspace}
\usepackage[hyphens]{url}
\usepackage{circledtext}
\usepackage{amsmath}
\usepackage{booktabs} 

\usepackage{hyperref}
\hypersetup{  
    breaklinks=true,
    colorlinks=true,
    urlcolor=blue
}

\newcommand{\ignore}[1]{}
\newcommand{\bm}{\texttt{Llama-3.1-8B }}
\newcommand{\ours}[0]{RAPID\xspace}
\newcommand{\longbench}[0]{LongBench\xspace}

\usepackage{array}
\newcolumntype{C}[1]{>{\centering\let\newline\\\arraybackslash\hspace{0pt}}p{#1}}
\usepackage{array}
\newcolumntype{M}[1]{>{\centering\arraybackslash}m{#1}}
\usepackage{tikz}
\usepackage{tikz}

\newcommand{\circled}[1]{%
  \tikz[baseline=(char.base)]{
    \node[
      shape=circle,
      fill=black,
      text=white,
      inner sep=0.4pt,
      minimum size=0.8em,
      font=\normalsize\bfseries
    ] (char) {#1};
  }%
}

\usepackage[accepted]{mlsys2025style/mlsys2025}

\mlsystitlerunning{Power Aware Dynamic Reallocation For Inference}

\begin{document}

\twocolumn[
\mlsystitle{Power Aware Dynamic Reallocation For Inference}



\mlsyssetsymbol{equal}{*}

\begin{mlsysauthorlist}
\mlsysauthor{Yiwei Jiang}{equal,amd-rad,uw-madison}
\mlsysauthor{Sangeeta Chowdhary}{equal,amd-rad}
\mlsysauthor{Nathaniel Morris}{amd-rad}
\mlsysauthor{Rutwik Jain}{amd-rad,uw-madison}
\mlsysauthor{Srilatha Manne}{amd-rad}
\mlsysauthor{Sam Bayliss}{amd-rad}
\end{mlsysauthorlist}

\mlsysaffiliation{amd-rad}{AMD Research and Advanced Development}
\mlsysaffiliation{uw-madison}{Department of Computer Sciences, University of Wisconsin-Madison, Madison, Wisconsin, USA}

\mlsyscorrespondingauthor{Sangeeta Chowdhary}{sangeeta.chowdhary@amd.com}

\mlsyskeywords{Machine Learning, MLSys}

\vskip 0.3in
\begin{abstract}
Disaggregation has emerged as a powerful strategy for optimizing large language model (LLM) inference by separating compute-intensive prefill and memory-bound decode phases across specialized GPUs. This separation improves utilization and throughput under fixed hardware capacity. However, as model and cluster scales grow, power, rather than compute, has become the dominant limiter of overall performance and cost efficiency. In this paper, we propose \ours, a power-aware disaggregated inference framework that jointly manages GPU roles and power budgets to sustain goodput within strict power caps. \ours utilizes static and dynamic power reallocation in addition to GPU reallocation to improve performance under fixed power bounds. \ours improves overall performance and application consistency beyond what is achievable in current disaggregation solutions, resulting in up to a 2x improvement in SLO attainment at peak load when compared to a static assignment without an increase in complexity or cost.  
\end{abstract}
]
\printAffiliationsAndNotice{*Equal contribution} 

\section{Introduction}
The compound annual growth rate (CAGR) of generative AI revenue is astounding, with projections reaching a rate of nearly 40\% between 2024 and 2030~\cite{AIGrowth}. The growth rate covers not just the training of larger models, but the growth of inference for existing and new models which is expected to outpace training in resource usage and power~\cite{InferenceVsTraining, AllAboutInference}. Along with the high revenue growth comes significant power challenges. Data centers are projected to consume between 6.7\% to 12\% of total US power by 2028, a 52\% to 272\% increase since 2023~\cite{Berkeley_DC_2024}. Not only does this imply greater operational cost for the power consumed, but it also restricts where and how fast data centers can be built due to a limited power supply. The growth in AI runs counter to power grid limitations, leading to the primary limiter to AI capabilities and abundant intelligence being the availability of power rather than cost or other resources~\cite{sam_blog}. The metric of AI viability is now being expressed in terms of how much compute can one deliver for a fixed power budget or Compute/GigaWatts~\cite{amd-openai} .

Disaggregation has been proposed as a solution to improve throughput and increase resource utilization~\cite{DistServe}\cite{splitwise}. LLM disaggregation is a technique that separates the two main phases of large language model (LLM) inference – \emph{prefill} and \emph{decode} – and executes them on separate hardware pools.  The prefill and decode phases have distinctive and orthogonal characteristics with prefill being compute intensive and decode being memory intensive when operating with typical batch sizes. Disaggregation improves performance, efficiency, and resource utilization by reducing interference between the phases resulting from long prefill requests delaying decode stages. In addition, disaggregation enables each phase of LLM inference to be optimized and scaled independently based on the configuration of requests. 

Disaggregation also has implications on power allocation in power-constrained environments. The prefill and decode phases have divergent hardware and power requirements. Prefill is more compute-intensive and requires higher power than the decode phase which is memory-intensive. We can take advantage of the differences to combine disaggregation with power management by allocating more power to GPUs running prefill and less to GPUs running decode.  
We evaluate performance using goodput and SLO attainment (described in Section~\ref{sec:exp-setup}), which measure the number of requests completed within latency targets~\cite{DistServe}.
Figure~\ref{fig:Intro} below shows the results for \emph{goodput} as a function of  \emph{queries per second} (QPS) per GPU on an AMD Instinct\texttrademark{} MI300X GPU node with 8 GPUs running \bm with tensor parallelism of 1 (TP=1). All schemes utilize disaggregation with eight GPUs with a power cap of 4800W total GPU power. However, two of the schemes shift the number of GPUs allocated to each compute phase while our solution utilizes non-uniform power allocation to deliver the best overall results. 


\begin{figure}
    \centering
    \includegraphics[width=1\linewidth]{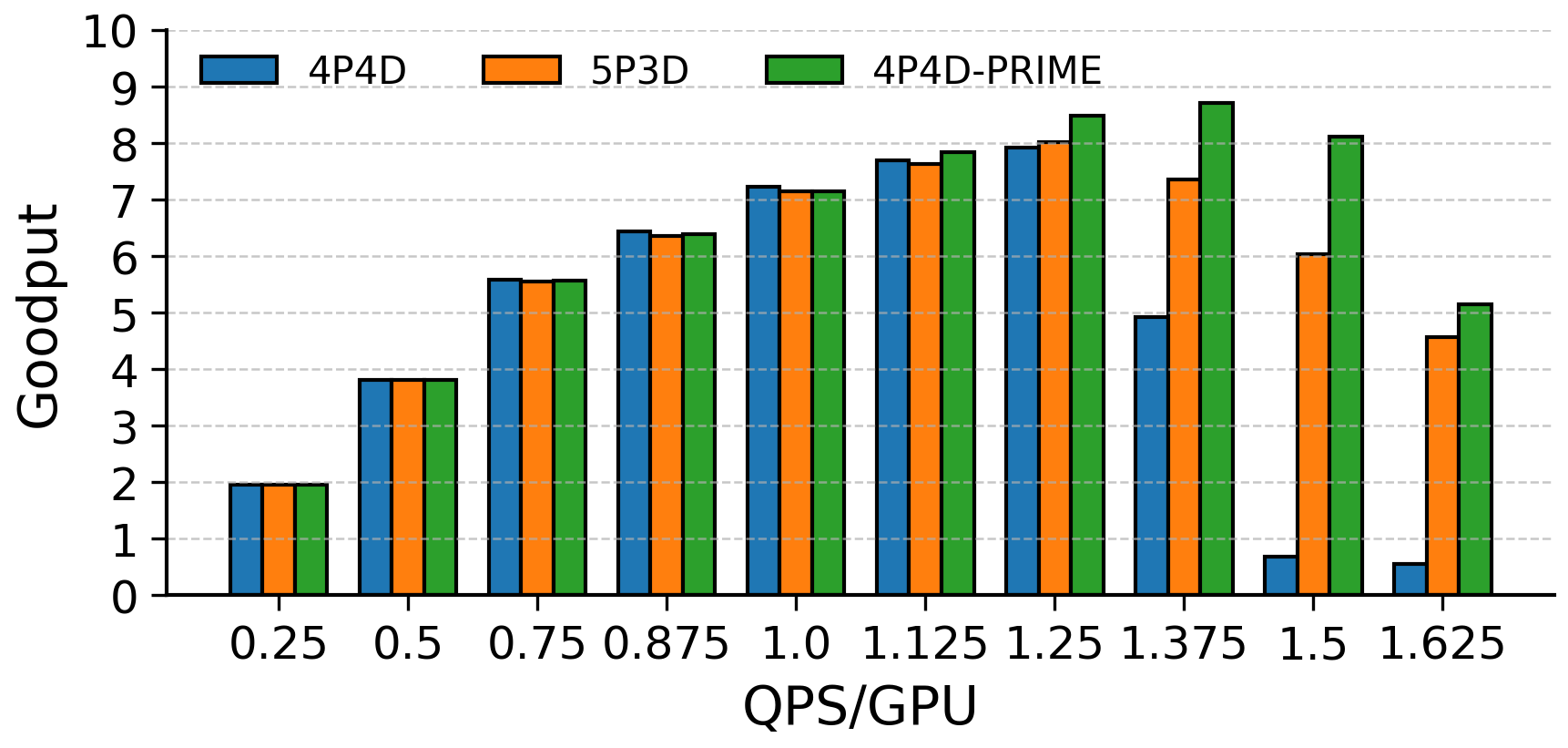}
    \caption{Goodput results for disaggregation schemes with varying GPUs (4 Prefill, 4 decode [4P4D], 5 prefill, 3 decode [5P3D], and 4 prefill, 4 decode with non-uniform power cap [4P4D]-\ours) as a function of queries per second (QPS) per GPU. All configurations use a 4800W node power budget. Higher goodput values are better.}
    \label{fig:Intro}
\end{figure}


In this paper, we present a resource allocation scheme called \ours that combines static and dynamic GPU allocation with power allocation to improve SLO attainment and Compute/W under power-constrained scenarios across disaggregation schemes. Past work has addressed related topics on disaggregation, dynamic GPU allocation, and/or power optimization in LLM inference. Table~\ref{tab:comparison} below compares and contrasts our paper to related works and is discussed further in Section~\ref{sec:related-works}. The key difference is that \ours is the only solution that combines GPU allocation with power allocation and SLO attainment to maximize performance in power-constrained scenarios and deliver the best \textit{Compute/W}.

\begin{table*}
    \caption{Comparison of existing work to \ours. }
    \centering
    \footnotesize
    \begin{tabular}{|C{6cm}|C{3cm}|C{2cm}|C{1.5cm}|C{2cm}|}

    \hline
        \textbf{Solution} & \textbf{Optimization} & \textbf{Dyn Power}  & \textbf{Disagg.} & \textbf{Dyn. GPU}  \\
                          &                       & \textbf{Alloc}      &                  & \textbf{Alloc}    \\
    \hline
        \textbf{POLCA}~\cite{LLM-ASPLOS} & Power Oversubscription            & No &    No  & No    \\
    \hline
        \textbf{DynamoLLM}~\cite{dynamollm} & Power Reduction & Yes        & No & No  \\  
    \hline
        \textbf{Splitwise}~\cite{splitwise} & Cost, Power & No        & Yes & Infrequent  \\  
                                                       & Oversubscription &     &  & \\
    \hline
        \textbf{DistServe}~\cite{DistServe} & Performance, TCO & No & Yes & No \\
    \hline
        \textbf{DynaServe}~\cite{DynaServe}, \textbf{WindServe}~\cite{WindServe} & Performance, TCO & No & Yes & Yes \\
    \hline
        \textbf{\ours} & Node Power Caps & Yes        & Yes & Yes  \\  
    \hline
    \end{tabular}
    \label{tab:comparison}
\end{table*}

Another unique aspect of this work is that our solution focuses on smaller models that run on a single GPU within a node rather than large models that run across a cluster.  \ours can be extrapolated to larger models using more GPUs and larger clusters, with power budgeting at different granularity -- either at the node level as the paper currently analyzes, or at the rack scale with a fully connected high-bandwidth interconnect between GPUs~\cite{amd_helios}. We focus on models that fit within a GPU for the following reasons: 1) the computational capabilities of a single GPU continues to increase, enabling ever larger models to fit within a single GPU~\cite{epoch2025nvidiachipproduction}; 2) large foundational models are not the best solution for many applications due to cost, privacy and security, and the complexity of adapting large models for the domain in question; and 3) smaller models offer tunability, flexibility, and cost savings and are expected to be deployed at a greater rates for enterprise and agentic use cases~\cite{slm_agentic, slm_enterprise}. The final reason for focusing on smaller deployments is the availability of power. Hyperscalars may be able to amass the power required for large-scale deployment in remote regions. However, enterprise customers require on-premises deployment close to metropolitan regions, which limits available power, especially in Europe and Asia, resulting in small models running on a few GPUs with throughput limited by immutable power constraints~\cite{edge_data_centers}.  

This work presents \ours. It combines disaggregation with static and dynamic GPU and power allocation. The goal is to improve performance within a fixed power budget. We measure performance using time-to-first-token (TTFT) and time-per-output-token (TPOT) SLO attainment. The system uses heterogeneous GPUs and power allocation together with disaggregation. This allows more compute to be provisioned within the same power envelope, improving overall compute per watt. The paper makes the following contributions: 
\begin{itemize}
    \item Proposes a disaggregation-conscious power allocation scheme to improve performance metrics; 
    \item Develops an algorithm that consolidates power constraints with the fluctuating requirements of incoming requests to deliver a performant solution for significantly less power; and 
    \item Incorporates power constraints into the vLLM scheduler for power-conscious disaggregation and presents results that show comparable performance for significantly less power. 
\end{itemize}

\section{LLMs and Power} 
GPU heterogeneity can be created through explicit control of power and frequency limits across GPUs within a node. Prior work has explored frequency scaling to improve the energy efficiency of inference~\cite{throttll'em, dynamollm}. In contrast, our approach focuses on adhering to power provisioning constraints to maximize compute capacity within a fixed power budget. The objective is to optimize goodput per watt while maintaining the required performance level.

\subsection{Power Capping}
Power capping each GPU within a node enables the provisioning of more systems within a fixed data center power budget.
Limiting GPU power has been considered in prior work~\cite{splitwise, LLM-ASPLOS}. The power limits, however, were set at the node or cluster level, and power was distributed equally across all GPUs. In this work, we examine capping power for GPUs within a node while distributing power unevenly across GPUs based on the characteristics of the function running on each GPU (prefill or decode) and SLO requirements. As long as the aggregate GPU power adheres to its power limit, the power allocated to each GPU can vary. 

\begin{figure}
    \centering
    \includegraphics[width=0.9\linewidth]{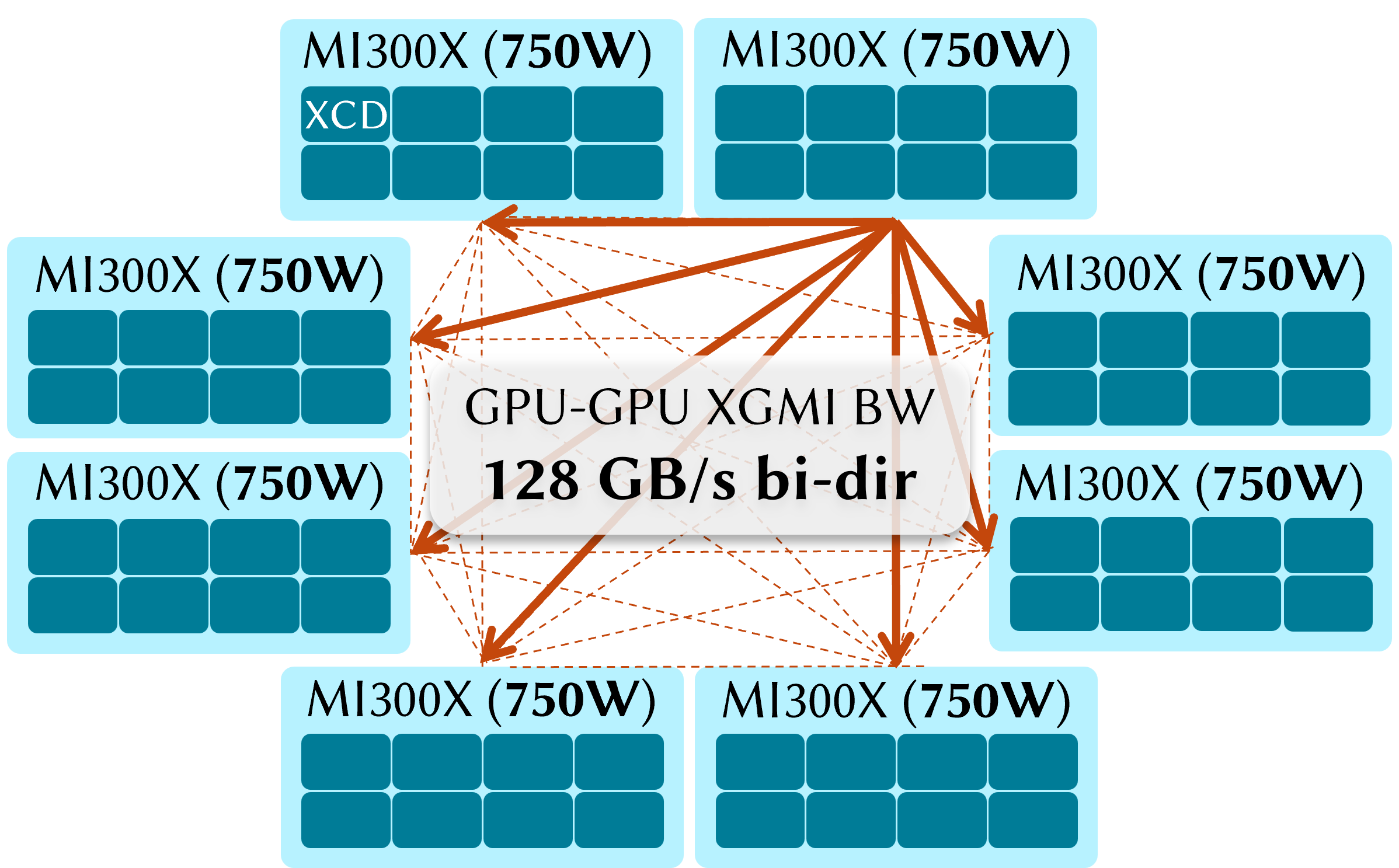}
    \caption{An eight GPU AMD Instinct\texttrademark{}~MI300X Platform showing XGMI bandwidth and GPU TDP limits.}
    \label{fig:mi300x}
\end{figure}

Consider the AMD Instinct\texttrademark{}~MI300X Platform (Figure~\ref{fig:mi300x}) with eight GPUs and a budget of 750W total board power (TBP) per GPU~\cite{MI300XPlatform}~\footnote{Total board power refers to the sum of all power consumption or losses on the GPU card including power delivery and voltage regulator inefficiencies, etc., and is an accurate representation of the total power consumed per GPU.}. Each GPU cannot sustain a higher power than its maximum rating, which is determined by the power delivery network and thermal limits. A fully provisioned AMD Instinct\texttrademark~MI300X GPU node requires $750 \times 8$ or 6000 watts to power all GPUs. GPU power can be lower than TBP when idle or running a light workload. Power oversubscription takes advantage of the difference between actual and provisioned power to add more devices to a fixed power budget, under the assumption that the system will not often exceed power limits~\cite {LLM-ASPLOS}. Power oversubscription requires mechanisms that monitor system power and throttle usage when power limits are at risk of being exceeded. When multiple GPUs draw high power simultaneously for sustained periods, frequent throttling can occur, reducing overall performance.

In contrast, \ours achieves higher compute capacity within the same power budget by managing disaggregated GPUs based on their phase characteristics. It can operate at the node level or, eventually, at rack scale, reducing the need for the complex power monitoring and control required in traditional oversubscription systems. Moreover, \ours can complement power oversubscription, optimizing performance within a node or rack under a fixed power envelope.

\begin{figure}
    \centering
    \includegraphics[width=1\linewidth]{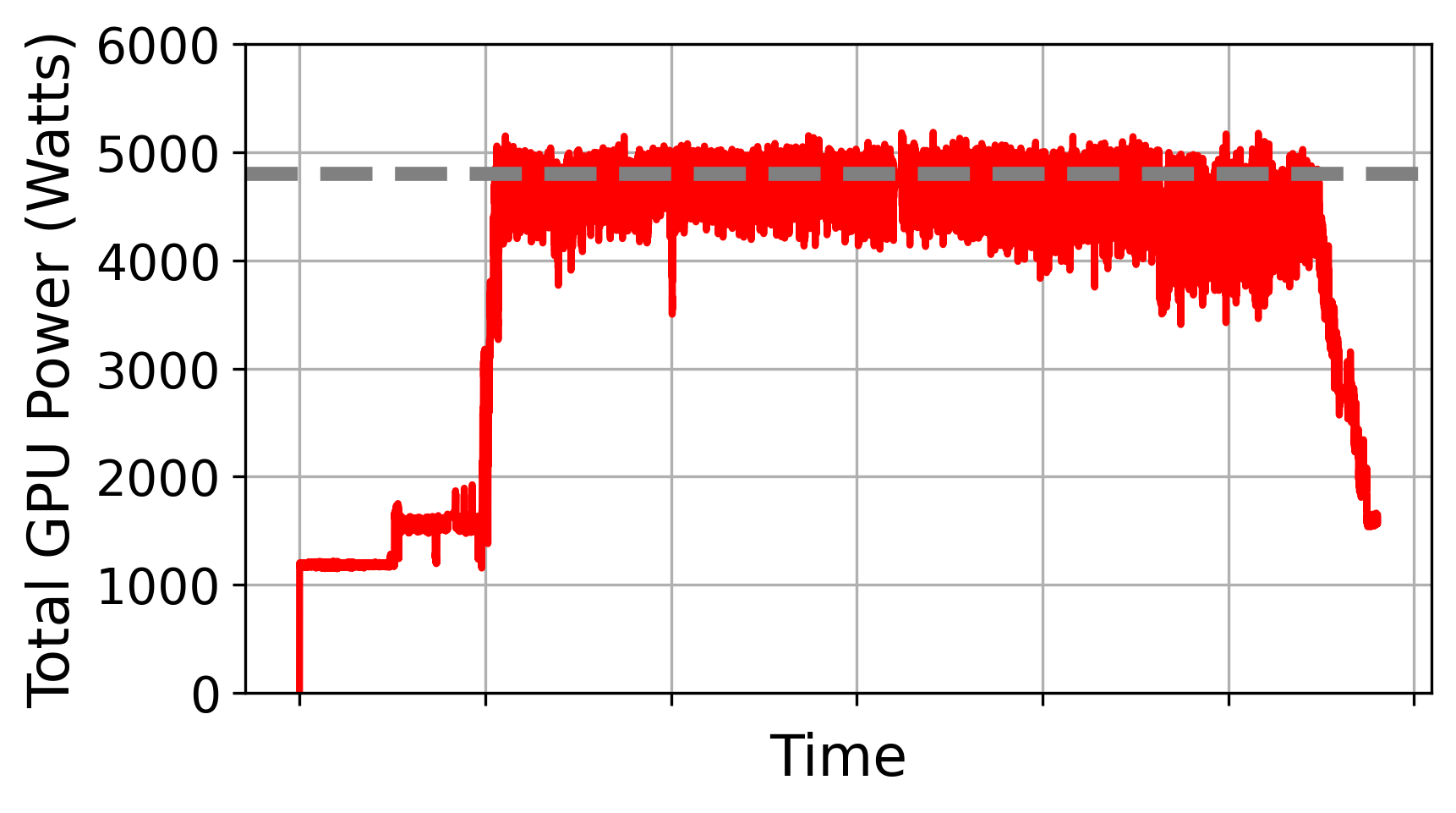}
    \caption{Time series of total GPU power for an uncapped node when running \longbench with a maximum input token size of 8K and QPS/GPU=1.5. Grey line indicates power limit of 4800W for total GPU power. }
    \label{fig:nodepowerlimits}
\end{figure}

In this paper, we evaluate with a node power budget of 4800 W. As shown in Figure~\ref{fig:nodepowerlimits}, the node frequently exceeds this limit. The figure illustrates a run of the \textit{LongBench} dataset with input sizes up to 8 K tokens and QPS / GPU = 1.5. Results are plotted as 10 ms rolling averages for the coalesced (non-disaggregated) configuration. Although total GPU power remains below the 6000 W hardware limit, there are many intervals where it surpasses the 4800 W budget.

Figure~\ref{fig:power-three}(a) and~\ref{fig:power-three}(b) below show the performance impact of power caps on the TTFT and TPOT for \bm with 4096 input tokens, 128 output tokens, varying batch sizes, and tensor parallelism of 1. Each GPU runs with power caps ranging from 400W to 750W in 50W increments. Each point in Figure~\ref{fig:power-three}(a) and ~\ref{fig:power-three}(b) represents the average across ten runs with a warm-up period between the runs to remove variability due to temperature. Performance results are relative to the 90th percentile latencies for the 400W configuration. The y-axis indicates the performance benefit of operating with a higher power budget. The \emph{prefill} phase, due to its computational intensity, is more sensitive to power caps than the \emph{decode} phase. Hence, the TTFT latency degrades more with lower power than the TPOT latency. TPOT continues to improve at a reasonable rate between 400W and 600W before flattening out, and TTFT begins to flatten out after 700W. Similar insights were provided in the Splitwise work~\cite{splitwise}, and they can be leveraged to distribute power unequally across GPUs in a disaggregated scenario based on how effectively each task meets its SLOs. 

\subsection{Dynamic Power Shifting}
\ours supports not only heterogeneous power allocation for GPUs within a node, but also dynamic power shifting between GPUs on the node. Power can be shifted between prefill and decode GPUs based on the workload requirement. How quickly power can be reallocated depends on how long it takes the power management firmware to reach the capped power limit. To avoid exceeding the total GPU power, \ours ensures that the \emph{source} GPU(s) whose power will be reallocated each have their power lowered before reallocating the power to the \emph{sink} GPU(s).

\begin{figure*}[t]
    \centering
    \begin{minipage}{0.32\textwidth}
        \centering
         \includegraphics[width=\linewidth]{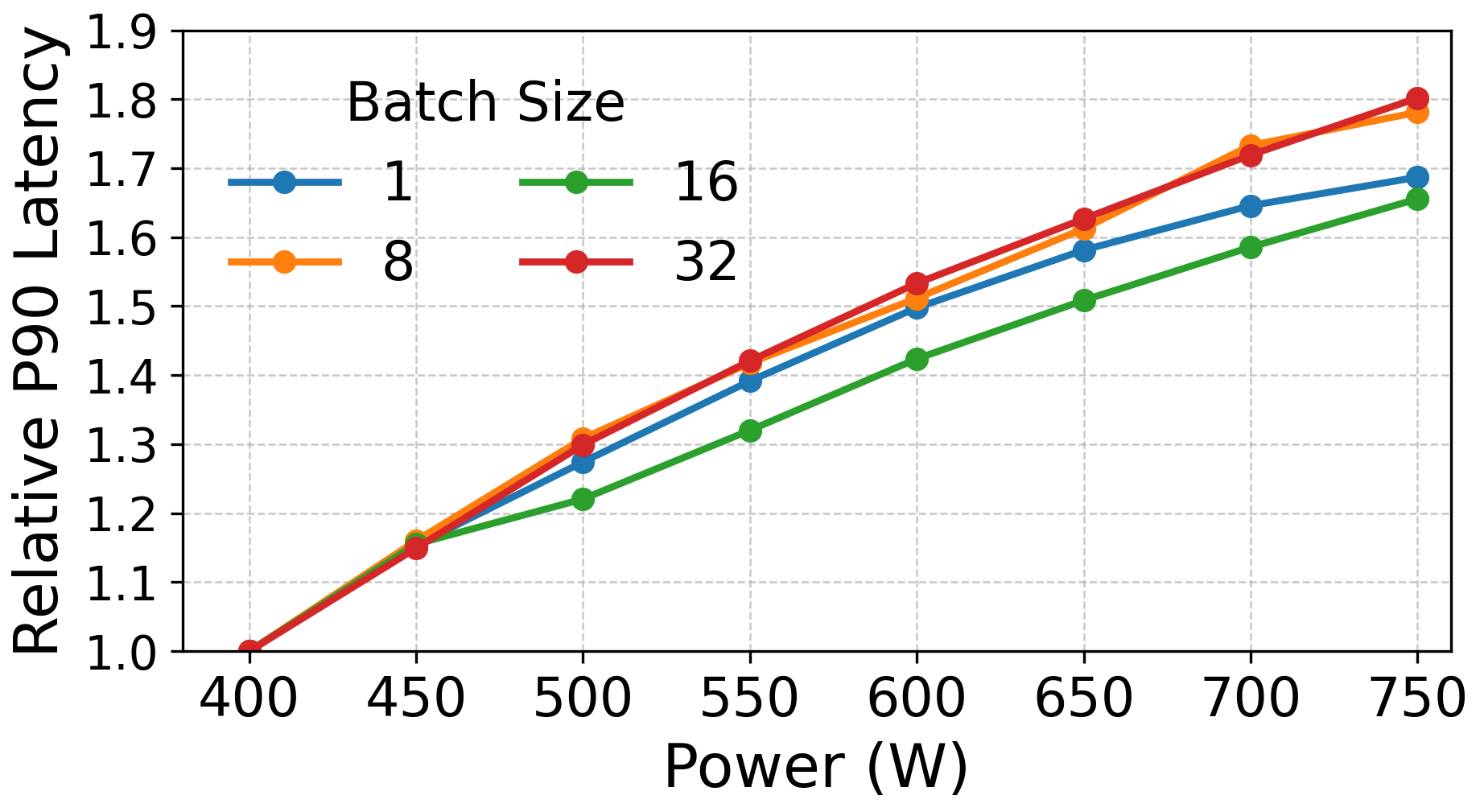}
        \vspace{0.3em}(a)
        \label{fig:prefill-power}
    \end{minipage}\hfill
    \begin{minipage}{0.32\textwidth}
        \centering
        \includegraphics[width=\linewidth]{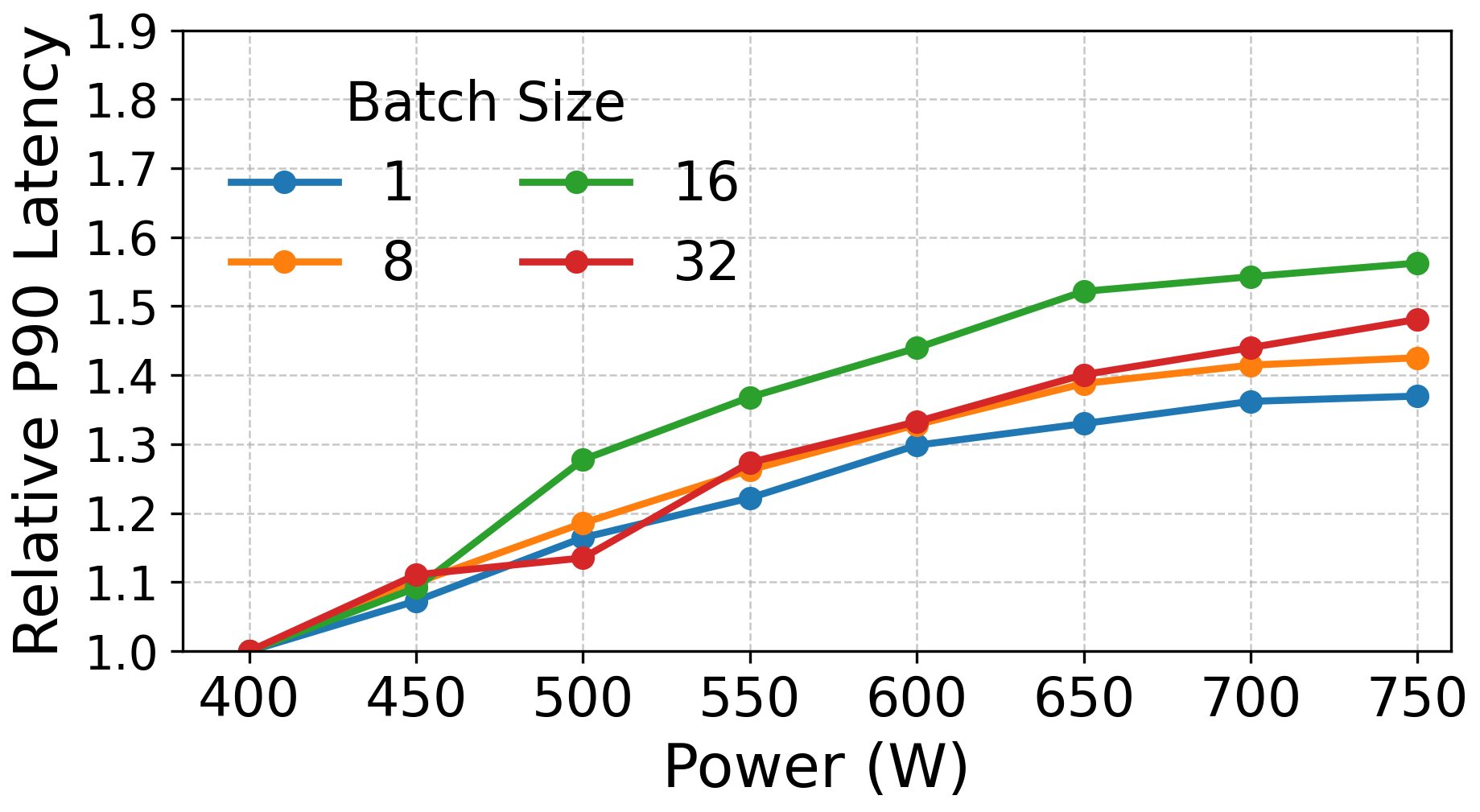}
        \vspace{0.3em}(b)
        \label{fig:decode-power}
    \end{minipage}\hfill
    \begin{minipage}{0.32\textwidth}
        \centering
        \includegraphics[width=\linewidth]{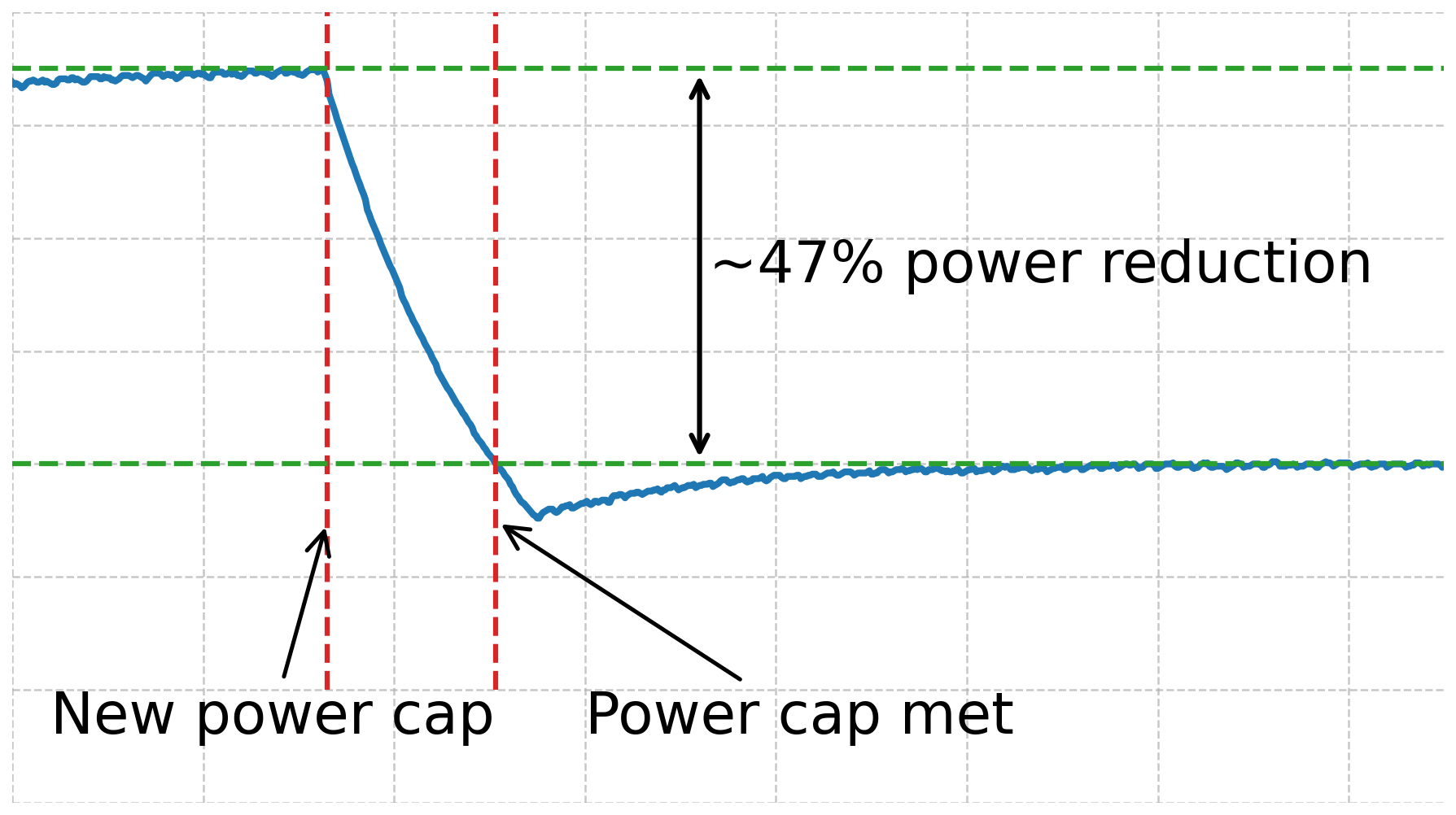}
        \vspace{0.3em}(c)
        \label{fig:power-cap}
    \end{minipage}

    \caption{(a) \emph{Prefill} P90 TTFT and (b) \emph{Decode} P90 TPOT latency as a function of GPU power and batch size. (c) Effectiveness of power cap initiated by \texttt{amd-smi} command.}
    \label{fig:power-three}
\end{figure*}

We analyze the power capping capabilities of the AMD Instinct\texttrademark~MI300X GPU using AMD SMI~\cite{amd-smi}. Other works have utilized frequency capping as a proxy for power capping due to the non-deterministic nature of power capping on other systems~\cite{LLM-ASPLOS, dynamollm}. Figure~\ref{fig:power-three}(c) shows the result of a 47\% reduction in power allocated to the GPU. The left side vertical line shows the initiation of the power cap, and the right side vertical line shows when the power cap was met. Due to the large reduction in the power cap, the power manager does not cap the power immediately, but requires some time to reach the new power limit. Once reached, the system adheres to the power limit. \ours uses a conservative value in the hundreds of milliseconds, from when a power-reduction command is sent to the source GPUs until the sink GPUs can increase their power budget. All of this is managed by vLLM through AMD SMI command-line interface~\cite{amd-smi}.

\section{Disaggregation and Power} 

Disaggregation refers to the splitting of phases for inference into two  groups: \emph{prefill} and \emph{decode}. The phases generally have orthogonal characteristics and splitting them reduces conflicts between the phases and enables load balancing each phase based on the request characteristics. In addition, it enables independent optimization of each phase either with specialized hardware~\cite{rubin-cpx} or by judicious power management with approaches like \ours. 
\begin{algorithm}[H]

\caption{Dynamic Resources Scheduling}
\label{alg:dynamic}
\begin{algorithmic}[1]
\State \textbf{Input:} GPU snapshot $(P, D)$, power budget $B$ \\
 \quad \quad \quad recent metrics $(\textit{TTFT}, \textit{TPOT}, \textit{rate}_P, \textit{rate}_D)$ \\ \quad \quad \quad queues $(Q_P, Q_D)$
\State \textbf{Constants:}$\mathrm{MIN_P}$,~$\mathrm{MAX_P}$,~$\mathrm{TTFT}_{\mathrm{SLO}}$,~$\mathrm{TPOT}_{\mathrm{SLO}}$
$\mathrm{THRESHOLD}$,~$\mathrm{MIN\_TIME}$,~$\mathrm{COOLDOWN}$
\State \textbf{Variables:} $last\_move\_time \gets 0$  \\
\quad \quad \quad \quad ~$now \gets \text{current time}$
\vspace{1mm}

\vspace{1mm}
\While{True}:
    \If{ $(\textit{TTFT} > \textit{TTFT}_{\text{SLO}})$
    $\land$ $(|Q_P| > \text{THRESHOLD})$ \\$\land$ $(\textit{TPOT} < \textit{TPOT}_{\text{SLO}})$ 
    $\land$ \\ $((now - last\_move\_time)>\text{COOLDOWN})$}
        \State \Call{MovePower}{from=Decode, to=Prefill}
       \If{\Call{PowerLimitsReached}{}}
            \State \Call{MoveGPU}{from=Decode, to=Prefill}
            \State \Call{DistributeUniformPower}{AllGPUs}
        \EndIf
        \State $last\_move\_time \gets now$
    \ElsIf{$(\textit{TPOT} > \textit{TPOT}_{\text{SLO}})$
    $\land$ $(\textit{TTFT} < \textit{TTFT}_{\text{SLO}})$ $\land$ $((now - last\_move\_time) > \text{COOLDOWN})$}
        \State \Call{MovePower}{from=Prefill, to=Decode}
        \If{\Call{PowerLimitsReached}{}}
            \State \Call{MoveGPU}{from=Prefill, to=Decode}
            \State \Call{DistributeUniformPower}{AllGPUs}
        \EndIf
        \State $last\_move\_time \gets now$
    \EndIf
     \State \texttt{sleep($\mathrm{MIN\_TIME}$)}
    \EndWhile
\end{algorithmic}
\end{algorithm}

Numerous papers address disaggregation and hardware utilization~\cite{splitwise, Dynamo, PodAttention, DistServe, DynaServe, HexGen, WindServe}, and some cover the combination of power and disaggregation.  For instance, in the \emph{splitwise} paper, prefill and decode are disaggregated across pools of nodes with the whole node servicing either prefill or decode~\cite{splitwise}. In addition, the pools of nodes may utilize different hardware resources such as lower power and lower cost A100s used for the less compute intensive decode phase and higher power and higher cost H100s used for prefill. 

Other works address dynamic resource reallocation to manage request rate and output token variability~\cite{DynaServe, WindServe}. Request rate is defined as the number of incoming queries per second. Request rate, along with the number of input and output tokens, directly governs the system’s utilization balance between prefill and decode GPUs. Unlike co-located setups, where both stages share the same resources, disaggregated architectures must continuously adapt to rate fluctuations to prevent pipeline stalls. High or bursty request rates increase prefill pressure, causing decode queues to fill faster than they can drain, whereas
sparse or variable arrivals can leave decode GPUs underutilized. This temporal variability, common in real-world workloads such as those from Microsoft~\cite{dynamollm}, challenges static resource allocation and motivates dynamic scheduling of GPU resources to prefill and decode. 

Disaggregation is an integral aspect of GPU management in production systems from both NVIDIA and AMD~\cite{Dynamo, AMD-blog}. Recently, specialized hardware designs have also been proposed to accelerate the prefill phase in disaggregated deployments~\cite{rubin-cpx}. Overall, disaggregation improves service throughput for a fixed set of resources by decoupling execution phases and allocating resources according to the specific demands of each phase. 

\subsection{Disaggregation Metrics}
LLM services have internal metrics that describe service level objectives (SLOs) or tiers that promise priority processing for customers and consistent speed even during peak demand~\cite{OpenAIPriority}. Meeting these SLOs is important to customer satisfaction, potentially requiring additional resources to meet SLOs during high demand phases. This not only increases cost, but also reduces the amount of compute achievable within a fixed power budget. 

Optimizing for throughput is not adequate for measuring performance under SLOs since it does not account for tail latencies and outliers. We use the \textit{goodput} metric described in ~\cite{DistServe} as the measure of goodness for this work. Goodput is defined as a measure of SLO attainment, focusing on both \textit{time to first token} (TTFT) and \textit{time per output token} (TPOT). Across a set of input requests, goodput tracks the number of requests that meet both the TTFT and TPOT SLOs utilized by the service provider. 

\subsection{\ours Implementation}
\label{sec:implementation}

\ours is implemented on top of the vLLM 0.8.4 framework for an AMD Instinct\texttrademark{}~MI300X GPU node~\cite{kwon2023efficient}. 
We expanded the existing vLLM implementation to execute across all GPUs within a node in a performant manner. 
Our implementation extends the vLLM engine to spawn separate worker processes, each bound to a dedicated GPU that can act as a prefill worker or a decode worker. A central scheduler process receives incoming requests, routes them to a specific worker, and coordinates inter-stage communication, i.e., the transfer of the KV-cache. Each worker process has a local scheduler that batches requests based on the GPU's memory capacity. 

We eliminate the host as a bottleneck by implementing direct GPU-to-GPU KV-cache transfer using HIP IPC and XGMI~\cite{hipcode, MI300XPlatform}. Memory pointers are shared across devices to achieve fast, contention-free communication at runtime. We implement a persistent ring buffer shared across GPUs for KV and hidden state transfer. When the KV-cache has been generated for all model layers, the prefill GPU publishes handles for the next available slot in the ring; the decode GPU can consume it as soon as the corresponding ready flag is set.  Each slot in the ring buffer contains the KV-cache per request, input token, and ROI metadata. To ensure correctness and minimize blocking, we use per-slot atomic ready flags and coordinate event-based synchronization with low-overhead polling.

Note that \ours utilizes bulk KV-cache transfer. Given the intra-node high-bandwidth connectivity, bulk KV-cache transfers did not incur significant latency overhead. Although we analyze intra-node disaggregation in this implementation, such a solution should scale to future rack-scale systems with tens to hundreds of GPUs, supported by low-latency communication fabrics to link the GPUs~\cite{amd_helios}. In addition, our disaggregation scheme uses a \emph{pull} mechanism for KV-cache transfers, with the decode GPUs retrieving data from the prefill GPU as needed while the prefill GPU continues with the next batch of inference requests. We maintain a request buffer of size 32, determined by the system's memory capacity.


\subsection{Dynamic Resource Management}
\ours combines dynamic GPU scheduling with dynamic power redistribution strategies to further optimize performance. Disaggregation reduces conflicts between prefill and decode and enables resource balancing based on request characteristics and the requirements of each phase. However, it treats all GPUs as being uniform. In \ours, we use power allocation in addition to dynamic GPU allocation to differentiate between prefill and decode GPUs and create heterogeneity across GPUs without resorting to bespoke hardware. The benefits of additional power, especially for the prefill phase, can be significant. According to Figure~\ref{fig:power-three}, a GPU running prefill can realize up to a 1.8x speedup for a 1.87x increase in power whereas decode speedup flattens out between 1.3x to 1.5x. By monitoring request arrival patterns and reacting to changes in rate and burstiness, \ours can reassign GPU roles and/or rebalance power budgets to maintain low tail latency and maximize compute/Watt efficiency.

We implement two types of allocation: static and dynamic. In the static allocation (\emph{Static}), the user statically assigns the number of prefill and decode GPUs, as well as the power allocated to each GPU. All GPUs assigned to the same phase have the same power allocation, but the power may differ between prefill GPUs and decode GPUs. In the dynamic solution (\textit{Dynamic}), the number of GPUs assigned to each phase and/or the power allocated to the GPUs in each phase can vary across execution. As in the static case, the power for the GPUs in each phase is homogeneous, but it can vary between decode and prefill GPUs. Shifting power from one GPU to another is relatively fast and requires minimal overhead. The only requirement is that power from the \emph{source} GPUs is reduced before being allocated to the \emph{sink} GPUs, so that the total GPU power for the node is not exceeded. 


The dynamic Algorithm~\ref{alg:dynamic} is reactive, and it allocates resources based on observed runtime behavior, including TTFT, TPOT, and input queue lengths. We manage power and  GPUs and move them to find a balance between prefill and decode workers. Rather than relying on latency prediction models or offline token-level profiling, our control loop monitors live queue pressure and recent performance issues to determine whether prefill or decode is currently limiting performance. When TTFT exceeds its target and decode GPUs are underutilized, power is shifted from decode to prefill or an idle decode GPU is reassigned to assist prefill. Conversely, if TPOT degrades and prefill appears idle, the system incrementally moves resources back to decode. This power shifting loop operates at sub-second intervals, following the requirements stated above to ensure node power limits are not violated. Shifting GPUs between phases incurs a larger performance overhead, requiring the GPUs to drain their current state before being shifting to the other phase. Hence, GPU reallocation occurs at a slower pace, on the order of two to five seconds. 

In contrast to WindServe~\cite{WindServe}, which uses model-driven simulations and latency estimations to predict future contention, our approach is fully observation-driven and requires no profiling or workload forecasting. While WindServe schedules requests based on predicted compute-to-communication ratios and prefetches GPU resources in anticipation of future demand, our design reacts to actual queue growth and SLO violations, offering faster adaptation and greater robustness in the face of noisy or unpredictable prompt distributions. Our reactive scheduler improves queue latency and SLO attainment under tight power budgets and disaggregated execution environments. 

To handle backpressure, the algorithm continuously monitors per-GPU input queue lengths as a proxy for overload in each role. When prefill GPUs show consistently large queues while decode queues are empty or lightly loaded, this indicates structural imbalance and triggers both power reallocation and GPU role reassignment to alleviate congestion. Unlike systems that rely solely on performance metrics like latency or throughput, we explicitly treat queue buildup as an early indicator of stress, allowing the scheduler to react before SLO violations become severe. 

To avoid oscillatory behavior caused by frequent power shifts or GPU role shifting, the scheduler incorporates a cooldown period between successive reallocation decisions. After any power adjustment or GPU migration, the control loop enforces a configurable delay (e.g., 2s–6s) before making additional changes, giving the system time to stabilize and absorb the impact of recent actions. This mechanism acts as a form of implicit hysteresis, ensuring that transient fluctuations in queue length or latency do not result in constant role-switching or ping-ponging of power between GPUs. The cooldown also prevents amplification of noise in reactive metrics like TTFT and TPOT, which are sensitive to small changes in batch composition or request arrival bursts. By combining queue-based triggering with a cooldown period between adjustments, our algorithm stays responsive while avoiding unnecessary or rapid changes, ensuring stability even under bursty or unpredictable workloads.

\section{Experimental Setup}
\label{sec:exp-setup}
We use the AMD Instinct\texttrademark{}~MI300X GPU with 192GB of HBM memory per GPU in an 8-GPU node configuration shown in Figure~\ref{fig:mi300x}. Each GPU has a total board power (TBP) of 750 watts per GPU. We built \ours on top of the vLLM 0.8.4 code base\footnote{vLLM is copyrighted by the vLLM Team and is subject to the Apache v2 license. You can find vLLM here: https://github.com/vllm-project/vllm.}. For the baseline, we used vLLM in coalesced mode. The results for the non-disaggregated configurations are generated using chunked prefill~\cite{ChunkedPrefill} while the disaggregated solutions utilize the implementation described in Section~\ref{sec:implementation}.  

Results are generated using \longbench and Sonnet datasets ~\cite{vllm_software, longbenchv2} with varying request rate. The \longbench dataset shows a unique distribution of long requests, which can affect the effectiveness of disaggregation and power allocation schemes. We limited the \longbench data to a maximum of 8K input tokens to fit reasonably within the experimental system and the SLO constraints set for the analysis.   
Sonnet is used to verify the robustness of the dynamic \ours algorithm for varying input sizes and distributions in a controlled manner. We use input token lengths of 8K and 512, and output token lengths of 128 and 512. We use a Poisson distribution for input arrival times.


We use \bm as the exemplar model for our analysis. We focus on a  
single-GPU model because GPUs are growing in size and capabilities, enabling ever-larger models to fit into a single GPU. At the same time, inference costs and the rise of enterprise and agentic use cases are expected to drive the adoption of smaller models. The AMD Instinct\texttrademark{}~MI300X GPU has 196GB of HBM capacity, while the upcoming MI450X GPU has 2.2x the memory capacity at 432GB~\cite{amd_helios} and can easily fit current mid-size or larger models. We focus on dynamic management within a single node for the paper, but these same principles and algorithms apply to multi-GPU models and rack-scale systems such as the Helios system with 72 GPUs~\cite{amd_helios}.  

The key metrics reported are TTFT and TPOT in the context of SLO attainment and goodput. TTFT is defined as the time required to process the prompt and generate the first token, and TPOT is the average time per token to generate the remaining output tokens~\cite{llm_benchmarking}. Because the decode GPU \emph{pulls} the KV-cache from the prefill GPU, any KV-cache transfer latency after the generation of the prefill token is included in the cost of the subsequent token generation. Hence the KV-cache transfer latency impacts TPOT and not TTFT. We also assess the compute capability per provisioned power using QPS/Watt as a metric of goodness that incorporates both performance improvements and provisioned power constraints. We analyze QPS/Watt using average provisioned GPU power. GPUs are the largest component of node power, accounting for 50\%-60\% of the node power budget with cooling, power delivery losses, and miscellaneous components (CPUs, NIC, DIMMs) accounting for the rest~\cite{LLM-ASPLOS}.

\section{Results and Analysis}
\label{results}
This section shows results for \ours with static and dynamic resource allocation using non-disaggregated (coalesced) and disaggregated solutions in a power-constrained budget of 4800W (600W/GPU). 

\subsection{Static Analysis}
\begin{figure*}[t]
    \centering
    \begin{minipage}{0.48\textwidth}
        \centering
        \includegraphics[width=\linewidth]{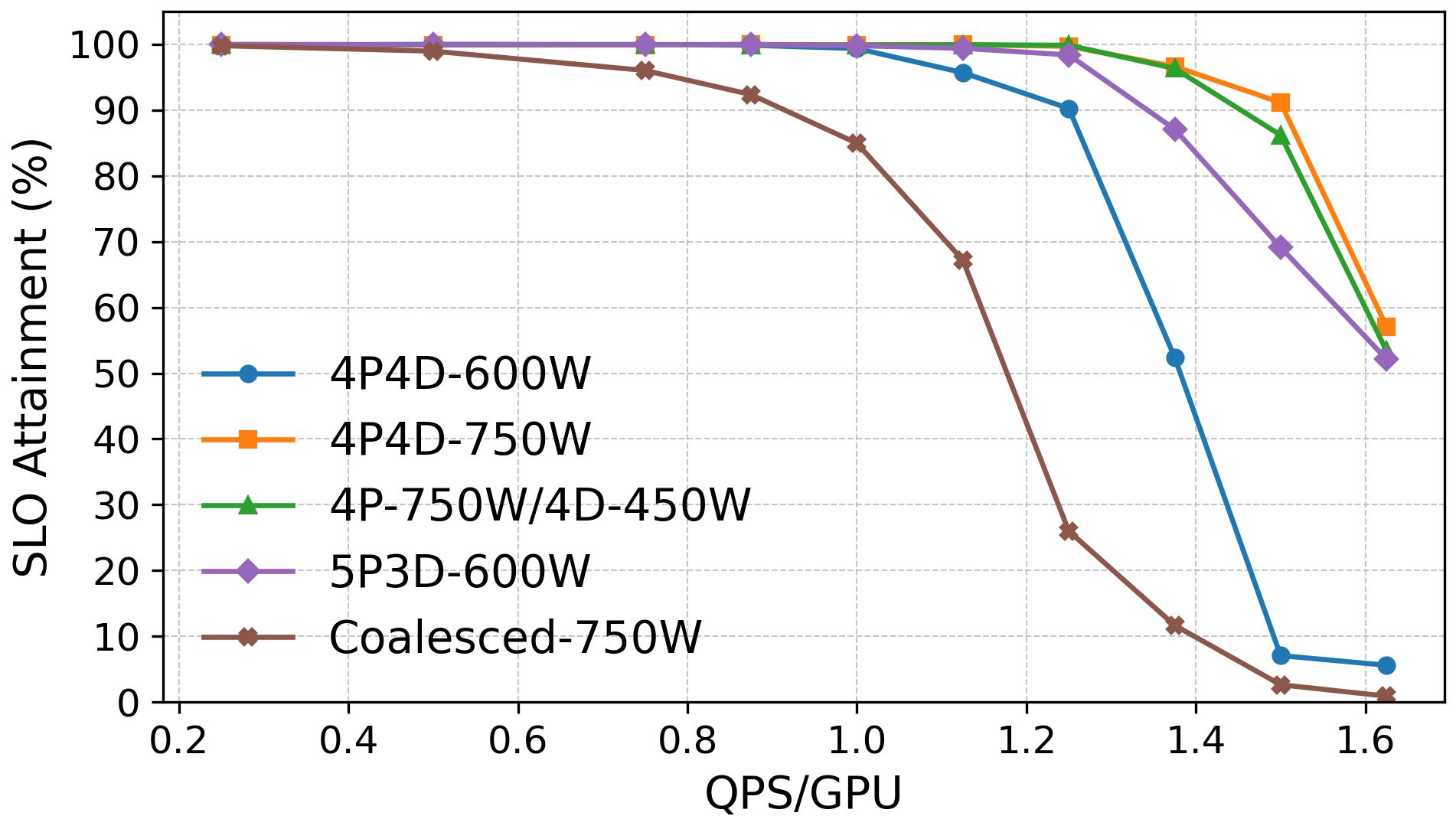}
        \vspace{0.3em}(a)
    \end{minipage}\hfill
    \begin{minipage}{0.48\textwidth}
        \centering
        \includegraphics[width=\linewidth]{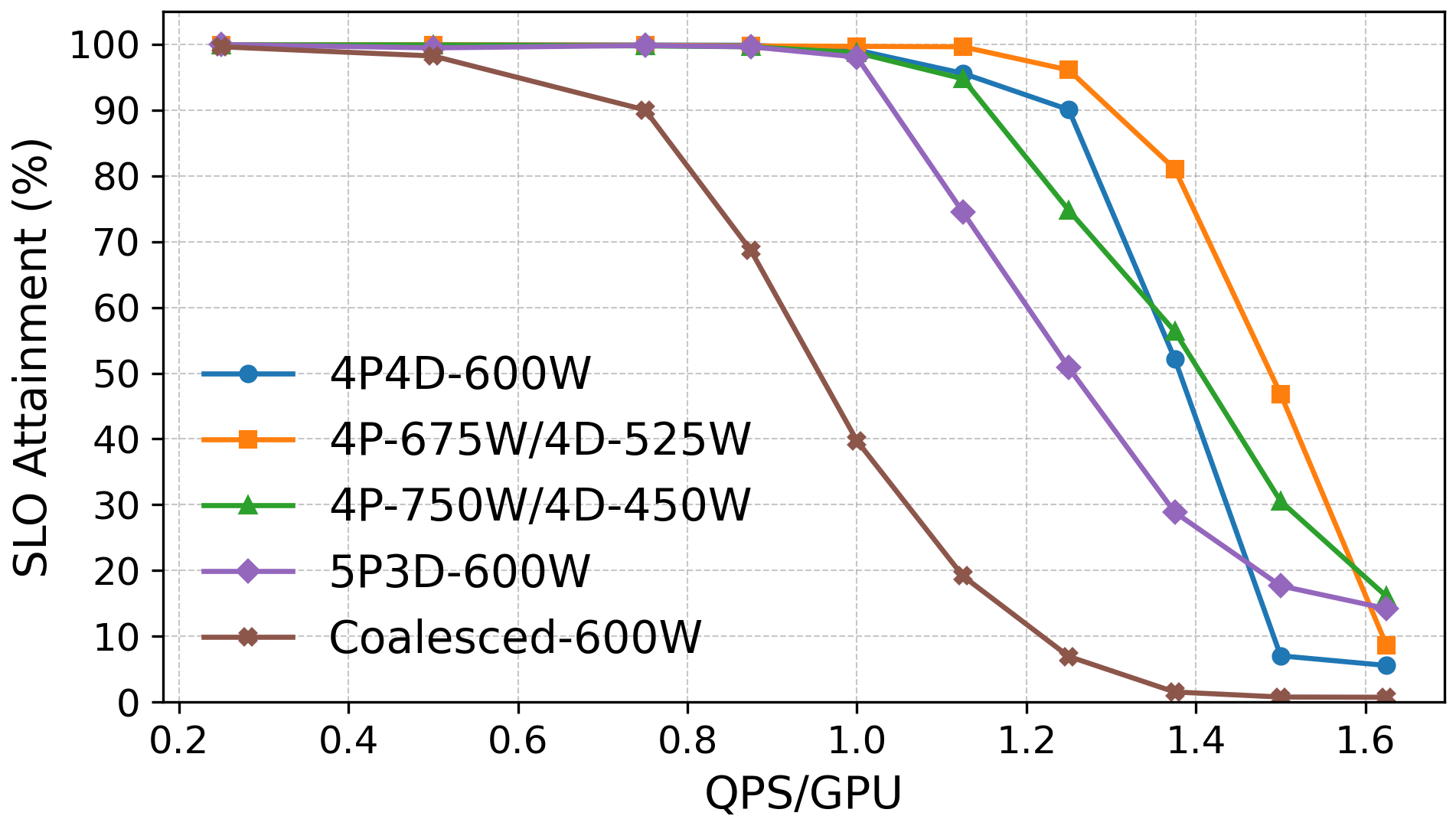}
        \vspace{0.3em}(b)
    \end{minipage}

    \vspace{0.3em}
    \caption{(a) SLO attainment with LongBench TTFT = 1 s and TPOT = 40 ms, 
             and (b) SLO attainment with LongBench TTFT = 1 s and TPOT = 25 ms.}
    \label{fig:static_long}
\end{figure*}
We use the \longbench dataset with up to 8K input tokens and SLOs of TTFT = 1 second, TPOT = 40ms, and 25ms to showcase the benefits of \ours. We select stricter SLOs compared to WindServe~\cite{WindServe} as we evaluate on a smaller model. We also show how our static approach scales with \longbench as the SLO scales for the fixed request rate. Reflecting the robustness of our approach as the latency requirements change, similar to DistServe~\cite{DistServe}. To understand the impact of power on performance, we ran the coalesced and disaggregated solutions at 6000W and 4800W. The best GPU and power allocations for the disaggregated solutions were determined empirically. We shifted GPUs between prefill and decode by increments of one, and shifted power by 50W in the 4P4D configurations to identify 4P-750W/4D-450W as the optimal configuration for 4800W solution.
\begin{figure}
    \centering
     \includegraphics[width=1\linewidth]{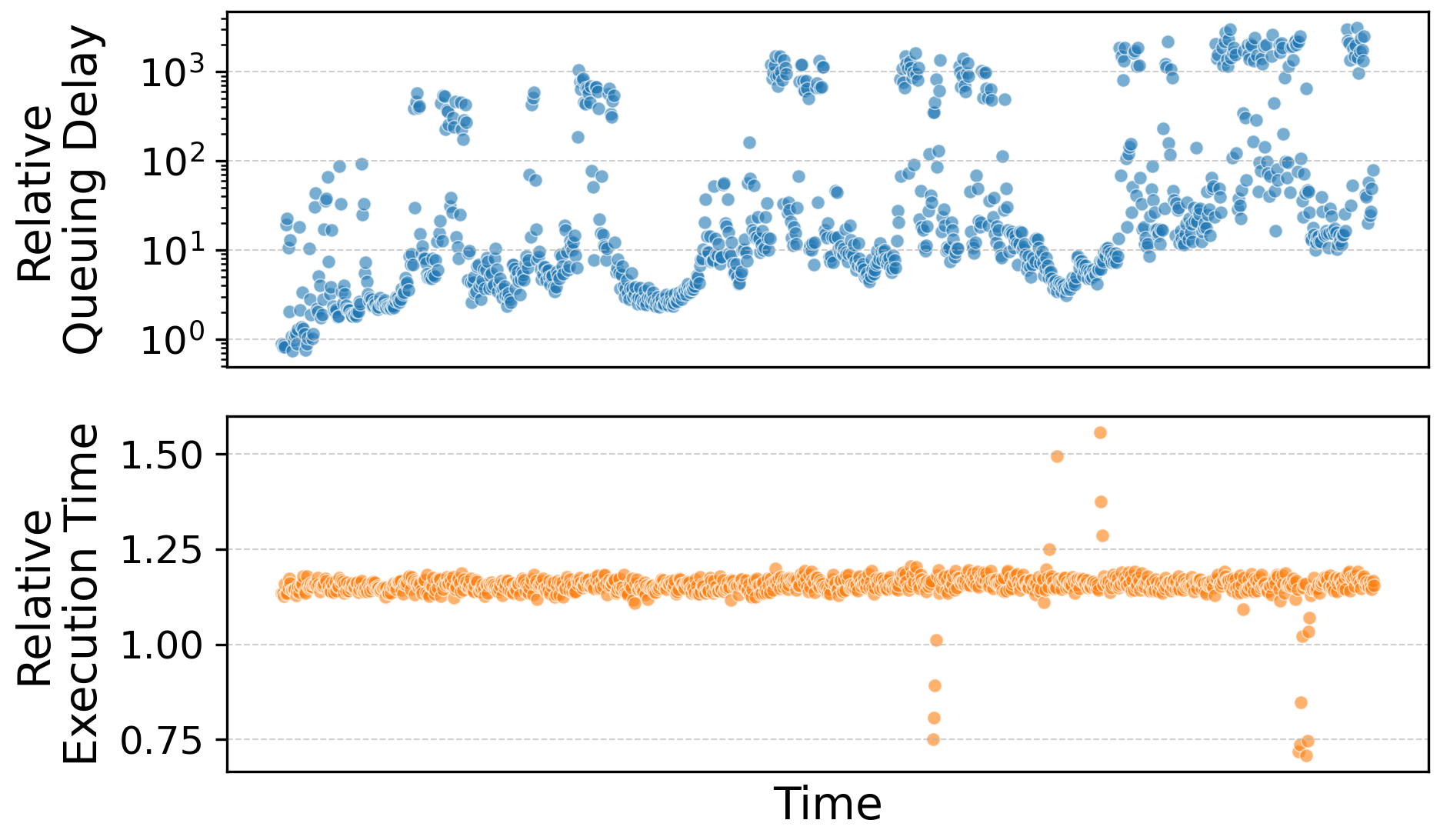}
    \caption{Queuing delay and execution time for 4P4D-600W relative to the 4P-750W/4D-450W configuration for \longbench. }
    \label{fig:RelTTFTLong}
\end{figure}

Figure~\ref{fig:static_long} (a) shows SLO attainment as the per-GPU request rate increases for \longbench with SLOs as TTFT = 1 second and TPOT = 40ms. At an 80\% SLO threshold, the 6000W disaggregated configuration with four prefill and four decode GPUs (4P4D-750W) sustains 1.5× higher request rate than the coalesced baseline (Coalesced-750 W). Dropping the power to 4800W (4P4D-600W) reduces the sustainable request rate to 1.2x that of the coalesced solution. However, with a 1200W reduction in GPU power and assuming GPU power accounts for 60\% of total node power, the resulting QPS/W is 1.36x higher than the coalesced 6000W solution. Finally, we evaluate a non-uniform power distribution (4P-750W/4D-450W) with prefill GPU power set to 750W and decode GPU power set to 450W. The 4P-750/4D-450 solution delivers comparable performance to a disaggregated solution (4P4D-750W) while consuming 1200 fewer watts. At 80\% SLO attainment, the configuration achieves 1.1× higher QPS/W than the disaggregated baseline (4P4D-750W) and 1.7× higher QPS/W than the coalesced 6000W baseline (Coalesced-750W). As a final option, we also evaluate a modified disaggregation scheme with five prefill and three decode GPUs at a similar power budget (5P3D-600W). Although this resulted in better performance than the 4P4D-600W solution, it did not achieve the same performance as the 4P-750W/4D450-450W solution. Shifting power from decode to prefill GPUs produced better results than adding more GPUs to the problem. 
 
To provide a deeper understanding of why SLO attainment improves with non-uniform power distribution, we analyzed  TTFT latency for the 4P4D-600W and the 4P-750W/4D-450W configurations at a rate of 1.5 QPS/GPU (Figure~\ref{fig:RelTTFTLong}). The latency is split into time spent in the queue (Queuing Delay) and time spent processing input tokens (ExecTime). The relative execution time holds steady across the run and is, on average, 15\% slower than the 750W prefill GPU in the non-uniform solution. However, the additional prefill latency accumulates, creating backpressure and increasing queuing delay. Queuing delay remains mostly negligible for 4P-750W/4D-450W but increases dramatically for the uniform power distribution case as the backpressure builds as requests are processed. 

Figure~\ref{fig:static_long}(b) shows SLO attainment for \textit{LongBench} when the TPOT target is tightened to 25 ms. Under these stricter SLOs, the performance of the 4P-750W/4D-450W configuration degrades, as allocating more power to prefill reduces the power available for decode and negatively impacts TPOT. In this setting, the 4P-675W/4D-525W configuration outperforms all other setups. This sensitivity to workload-specific SLOs underscores the importance of dynamic power allocation, in which power can be continuously adjusted between the prefill and decode phases based on observed performance.

\begin{figure*}
    \centering
     \includegraphics[width=1\linewidth]{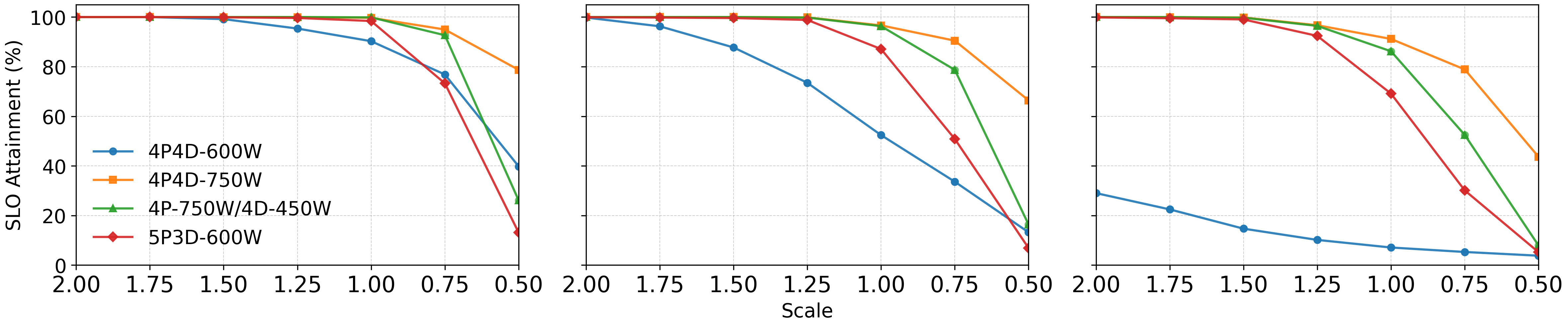}
    \caption{Comparison of SLO scaling for disaggregated solutions with uniform and non-uniform power allocation. 
    Each subplot corresponds to QPS/GPU = 1.25, 1.375, and 1.5, respectively.}
    \label{fig:scaling}
\end{figure*}

To examine how the system scales under changing latency requirements, we uniformly vary the SLOs from relaxed targets of TTFT = 2 s and TPOT = 80 ms (2× the baseline) to stricter targets of TTFT = 0.5 s and TPOT = 20 ms (0.5x the original SLOs). As shown in Figure~\ref{fig:scaling}, for QPS/GPU values of 1.25, 1.375, and 1.5, the non-uniform power configuration consistently outperforms other disaggregated setups under the same power budget and matches the performance of the higher-power 4P4D-750W configuration until the SLOs become highly restrictive.




\subsection{Dynamic \ours}
\label{sec:dyn-prime}
The analysis with varying SLOs highlights the need for a mechanism to identify the best configuration based on the characteristics of the input dataset and SLO requirements. Dynamic \ours achieves this goal by adjusting to the needs of the dataset and user requirements. 

To highlight the capabilities of dynamic \ours, we use Sonnet to create a synthetic workload with 1000 prefill heavy requests (8K input tokens, 128 output tokens) and 1000 decode heavy requests (500 input tokens, 500 output tokens) that follow a Poisson distribution for arrival times. We use the same TTFT SLO (TTFT=1sec) across the full workload, but tighten TPOT SLO from 40ms during the prefill heavy portion to 20ms for the decode heavy portion in order to stress both TTFT and TPOT during different phases of operation. 

 Figure~\ref{fig:dyn_prime} below shows SLO attainment with dynamic and static power and GPU management solutions.  The standard disaggregation solutions (4P4D-600W and 5P3D-600W) has the lowest SLO attainment. The static (4P-750W/4D-450W) and dynamic (4P4D-DynPower) power allocation schemes in \ours have similar SLO attainment because the 4P4D-DynPower solution converged on the same power distribution as the static 4P-750W/4D-450W solution. Neither of the power reallocation solutions achieve good results because power reallocation alone has limited benefit in decode heavy scenarios. The dynamic GPU allocation scheme (DynGPU-600W) allocates 600W to each GPU but is able to improve performance by shifting GPUs between prefill and decode  to balance workload requirements. Finally, the DynGPU-DynPower scheme dynamically allocates GPUs and power based on SLOs and performs best overall, achieving a higher QPS/GPU than all other configurations.  

\begin{figure}
    \centering
    \includegraphics[width=1\linewidth]{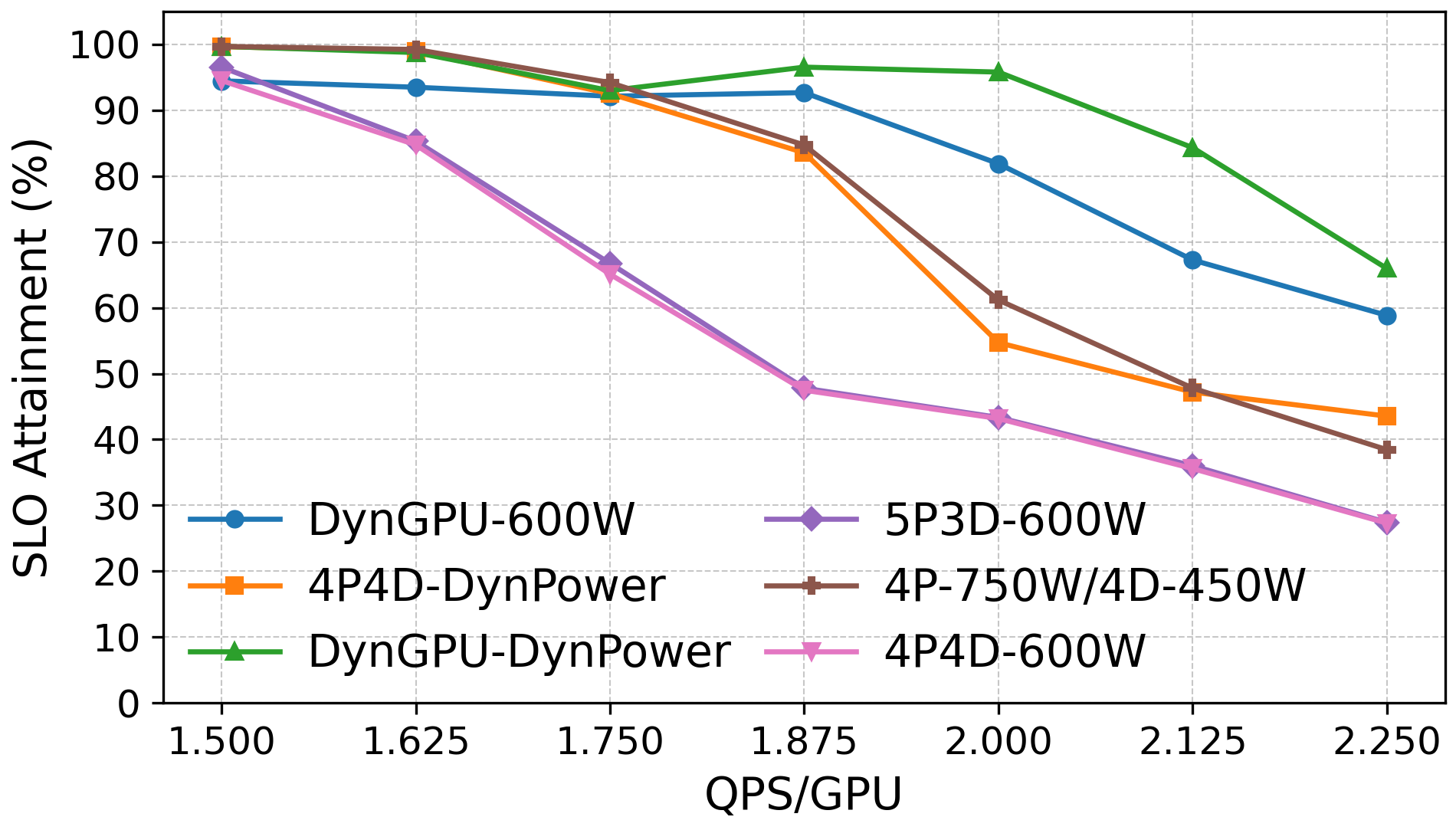}
   \caption{SLO attainment for static and dynamic \ours configurations. }
    \label{fig:dyn_prime}
\end{figure}

\begin{figure*}[t]
    \centering
    \begin{minipage}{0.32\textwidth}
        \centering
        \includegraphics[width=\linewidth]{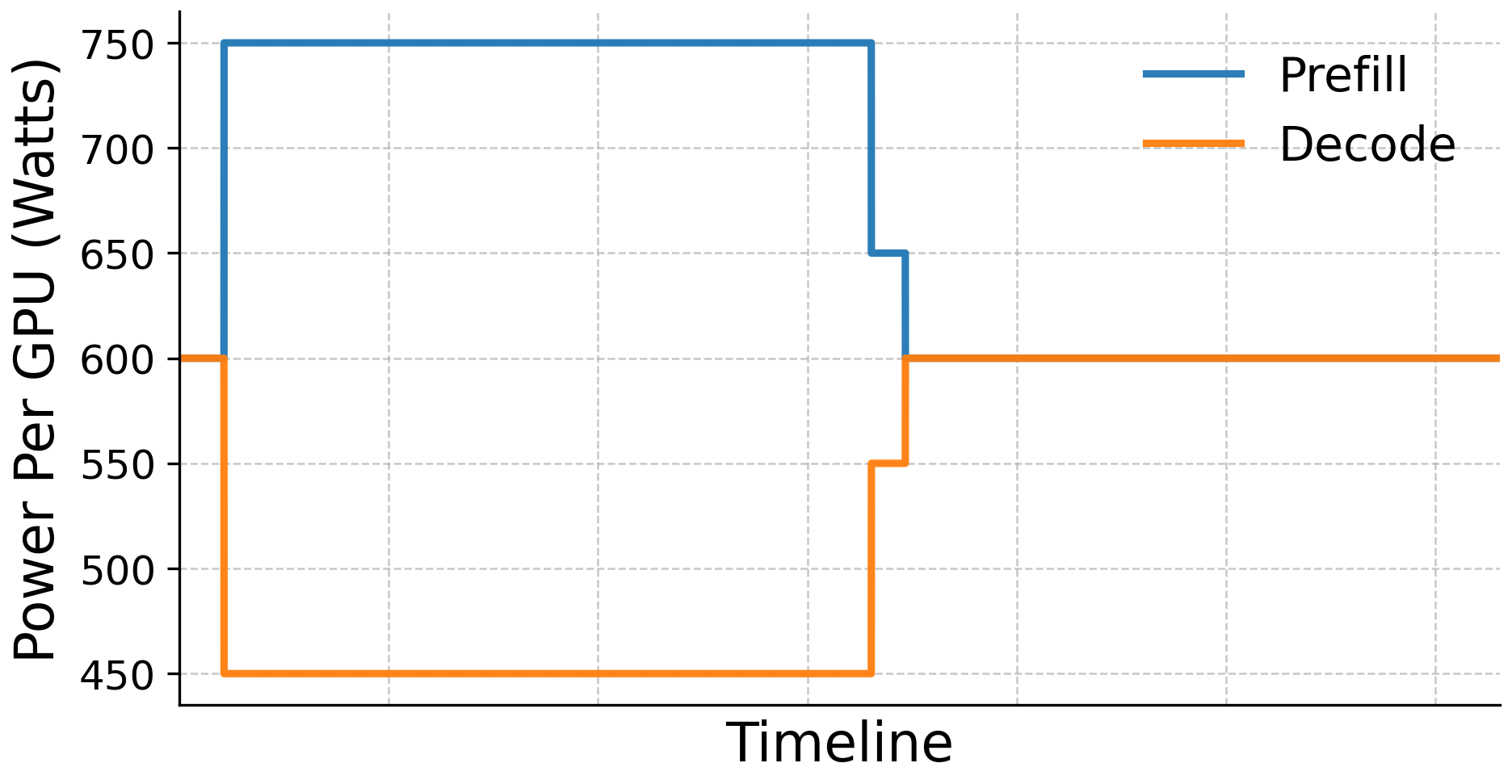}
        \vspace{0.3em}(a)
    \end{minipage}\hfill
    \begin{minipage}{0.32\textwidth}
        \centering
         \includegraphics[width=\linewidth]{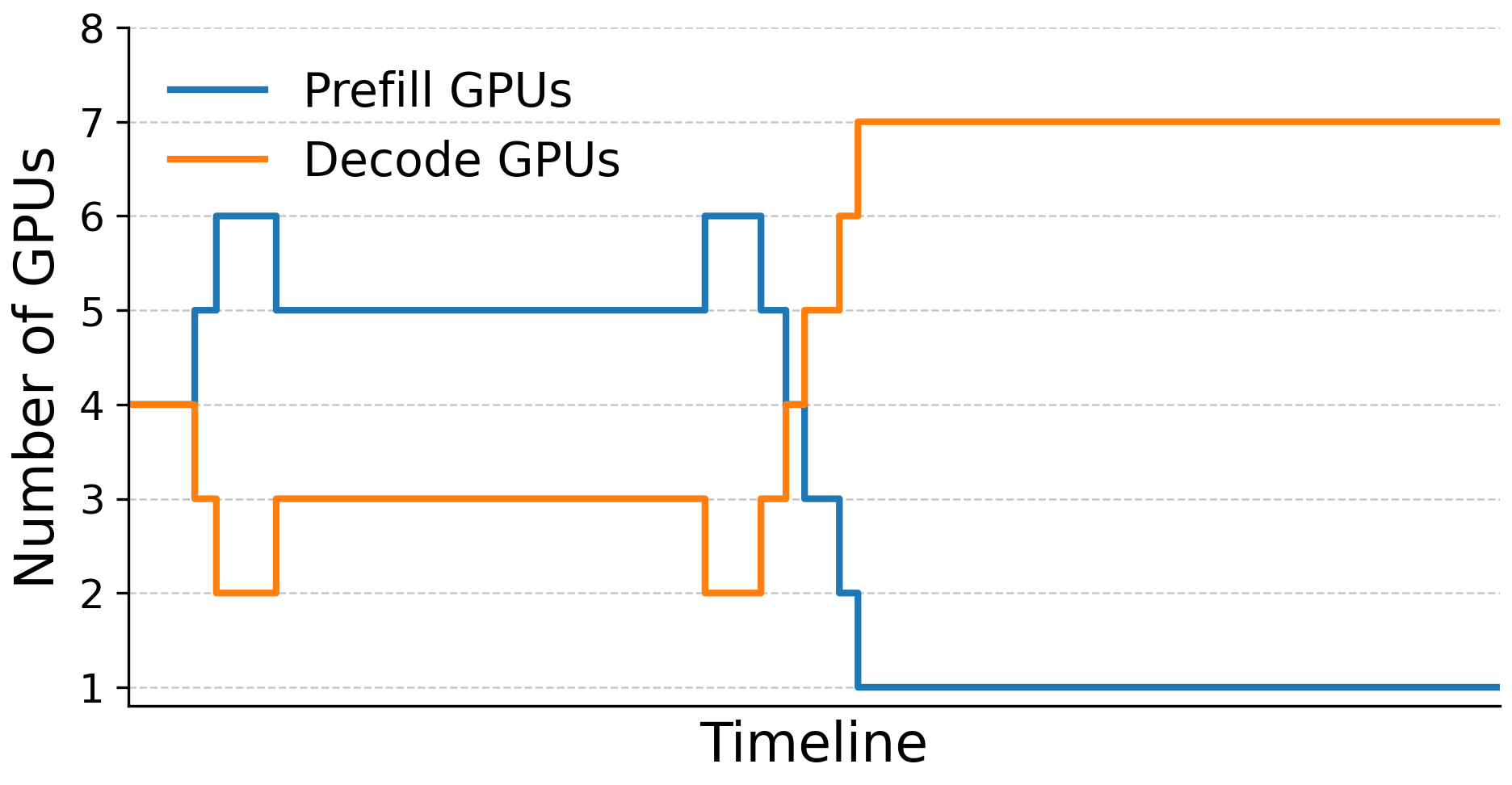}
        \vspace{0.3em}(b)
    \end{minipage}\hfill
    \begin{minipage}{0.32\textwidth}
        \centering
        \includegraphics[width=\linewidth]{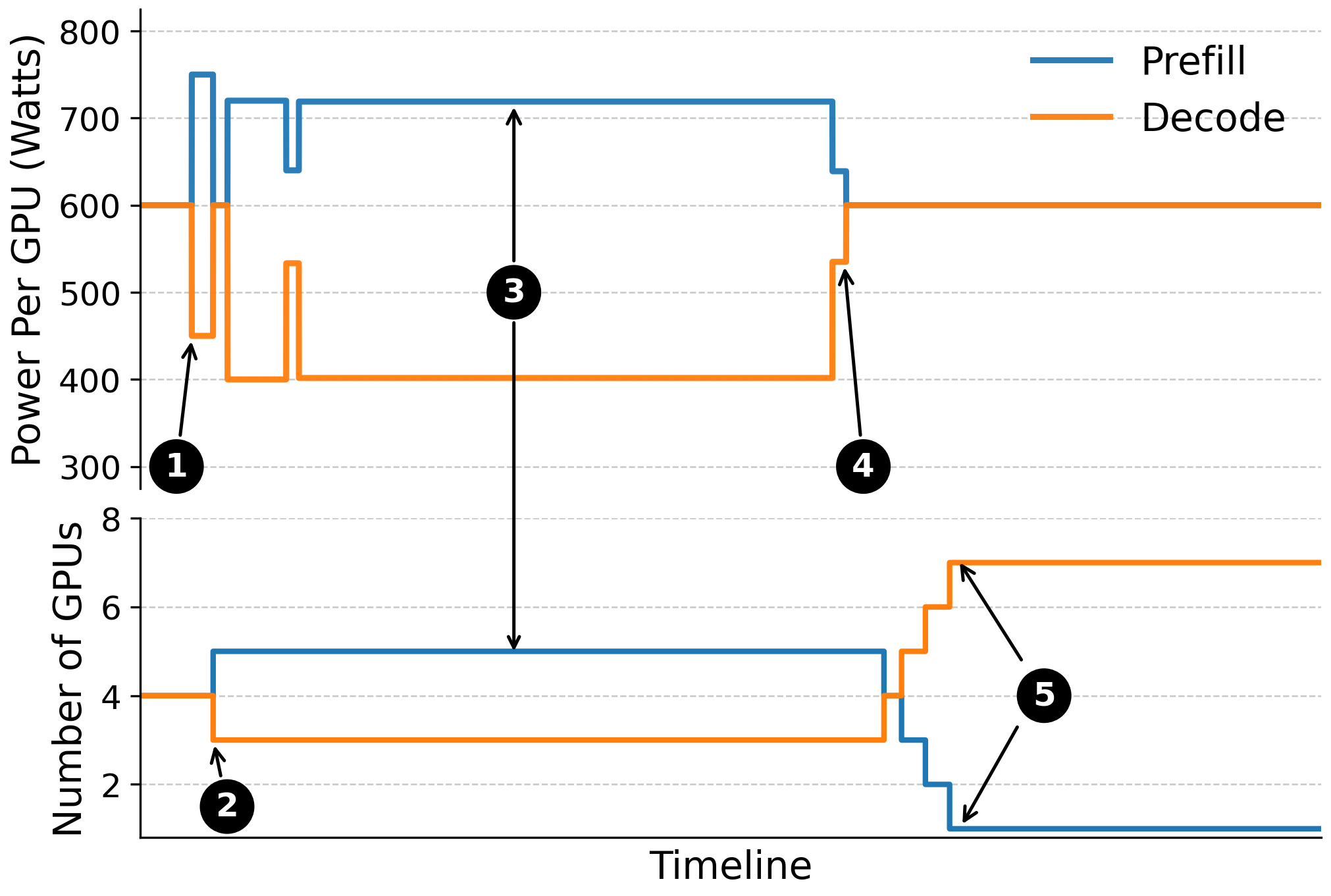}
        \vspace{0.3em}(c)
    \end{minipage}\hfill

    \vspace{0.3em}
    \caption{(a) Dynamic \ours with fixed GPUs and dynamic power allocation, (b) fixed power and dynamic GPU allocation, and (c) both dynamic GPU and dynamic power allocation.}
    \label{fig:dyn_gpu_comparison}
\end{figure*}

Figures~\ref{fig:dyn_gpu_comparison}(a) and (b) below illustrate dynamic \ours managing power and GPU allocation, respectively, for the Sonnet dataset results presented in Figure~\ref{fig:dyn_gpu_comparison} at a QPS/GPU rate of 2.0. The data collected for Figure~\ref{fig:dyn_gpu_comparison} below were averaged across three runs and the management analysis is from a representative run. During the prefill heavy portion, the 4P4D-DynPower solution maximizes prefill GPU power at 750W. For the decode heavy portion, the power across all GPUs is set to 600W. The algorithm limits decode GPUs to a peak of 600W because decode performance does not scale much above 600W (Figure~\ref{fig:scaling}). Prefill GPUs benefit from borrowing power from decode GPUs, but the reverse is generally not necessary. The DynGPU-600W configuration shows significant GPU shifting, dedicating up to six GPUs for prefill during the first phase and shifting to seven GPUs for decode during the second phase. The algorithm guarantees a minimum of at least one GPU for each phase. Dynamic GPU management delivers more compute cycles/s to the pool than modifying a GPU's power cap between 400W and 750W would. Hence \ours with dynamic GPU management achieves better results in scenarios with large variability across prefill and decode heavy phases. 

Figure~\ref{fig:dyn_gpu_comparison}(c) below shows how \ours combines dynamic GPU and power allocation to produce superior results. Starting with an even distribution in power and GPUs, \ours attempts to meet SLOs by moving power to prefill GPUs at \circled{1}. When this is not sufficient, \ours reallocates a decode GPU to prefill at \circled{2}. The combination of an additional GPU and additional power at \circled{3} satisfies SLOs attainment until the dataset begins to transition from the prefill-heavy portion to the decode-heavy portion at \circled{4}. At this point, power is reallocated to the decode GPUs. When this is not sufficient, GPUs are reallocated from prefill to decode until seven GPUs shift to decode to maintain SLOs. At the same time, power remains constant at 600W at \circled{5}, matching the behavior of the dynamic GPU allocation scheme in Figure~\ref{fig:dyn_gpu_comparison}(b). By combining dynamic power and dynamic GPU allocation, \ours improves upon both schemes and achieves the best overall results. 

To summarize, we showed the benefits of non-uniform power distribution between prefill and decode GPUs in a disaggregation scenario, highlighting how judicious power allocation across GPUs within a node can improve disaggregation results. We analyzed how a dynamic allocation mechanism can adapt GPU and power allocation to changing patterns and requirements. Finally, although the solutions presented were at the node level, we expect these solutions to translate to larger rack-scale systems with tens to hundreds of GPUs.

\section{Related Work}
\label{sec:related-works}

Disaggregated inference is gaining attention in both academia and industry for improving latency and hardware utilization in large language model (LLM) serving. Recent systems decouple prompt prefill and token-by-token decode across GPUs or nodes to avoid head-of-line blocking and improve throughput under bursty workloads \cite{splitwise, DistServe}. Prior work, such as Splitwise and DistServe, demonstrates that separating prefill and decode enables finer GPU specialization and load balancing. However, these systems assume homogeneous power provisioning and ignore asymmetric power constraints. We extend disaggregated inference by introducing power-aware disaggregation, wherein distinct power caps are applied to prefill and decode GPUs based on their workload profiles. Unlike previous efforts, our system treats power as a first-class resource alongside compute, using a dynamic scheduler to reallocate power between GPUs to meet service-level objectives (SLOs) and improve goodput.

\textbf{WindServe}~\cite{WindServe} introduces a stream-based token scheduler to reduce tail latency in phase-disaggregated serving but assumes uniform power provisioning. In contrast, our work applies differentiated power caps and adapts scheduling based on power telemetry, optimizing both latency and energy efficiency. \textbf{Splitwise}~\cite{splitwise} most closely resembles \ours, splitting prefill and decode across systems and allocating machines to prefill, decode, or mixed pools. However, resource reallocation in Splitwise is coarse-grained, and power budgets are static. Our system instead reallocates power dynamically between prefill and decode, or heterogeneously across prefill GPUs, at fine temporal granularity to exploit available power as request rates vary. Other systems~\cite{DistServe, DynaServe, WindServe} explore GPU reallocation and KV-cache transfer overheads but overlook power management.

\textbf{POLCA}~\cite{LLM-ASPLOS} manages cluster-level power via frequency caps to avoid power violations, requiring empirical tuning and recovery mechanisms. In contrast, \ours enforces GPU-level power caps directly, eliminating the need for empirical control. \textbf{DynamoLLM}~\cite{dynamollm} adjusts frequency and tensor parallelism for energy efficiency under SLOs but cannot guarantee a fixed power draw. Our system ensures explicit adherence to provisioned power budgets using hardware-enforced caps.

Prior schedulers, such as Clockwork \cite{258862} and Tiresias \cite{227623} target batch DNN workloads rather than token-level, stateful LLM inference. Our power-aware, stage-specific scheduling dynamically assigns prefill and decode roles per request based on power and load signals.

To our knowledge, this is the first system to combine static and dynamic asymmetric power capping with GPU-level disaggregation for LLM inference. By aligning power provisioning with workload demand, more power for prefill, less for decode, we achieve better latency and energy efficiency under tight power budgets. Prior disaggregated systems overlook power asymmetry, and prior power-aware systems lack stage-level disaggregation; our approach unifies both, demonstrating their synergy.

\section{Conclusions \& Future Work}
Disaggregated inference has emerged as an effective approach to reduce latency and improve GPU utilization in large language model (LLM) serving. 
We presented a power-aware, disaggregated inference architecture that leverages the asymmetric power requirements of prefill and decode to maintain QoS under tight performance constraints while improving latency robustness and energy efficiency. This work analyzes node-level power management, but the underlying algorithms can be applied to rack-scale deployments, larger models, and hardware with higher power-delivery capabilities. Future work involves expanding to multi-GPU ($TP>1$) model configurations and incorporating predictive methods into the dynamic mechanism for determining ideal \ours configurations.

\bibliography{ref}
\bibliographystyle{mlsys2025style/mlsys2025}



\end{document}